# Observed Metallization of Hydrogen Interpreted as a Band Structure Effect

Mehmet Dogan[1,2], Sehoon Oh[1,2], Marvin L. Cohen[1,2,*]

[1] Department of Physics, University of California, Berkeley, CA 94720, USA
[2] Materials Sciences Division, Lawrence Berkeley National Laboratory, Berkeley, CA 94720, USA
[*] To whom correspondence should be addressed: mlcohen@berkeley.edu

**Abstract**

A recent experimental study of the metallization of hydrogen tracked the direct band gap and vibron frequency via infrared measurements up to ~425 GPa [P. Loubeyre *et al.*, *Nature* **577**, 631 (2020)]. Above this pressure, the direct gap has a discontinuous drop to below the minimum experimentally accessible energy (~0.1 eV). The authors suggested that this observation is caused by a structural phase transition between the *C2/c*-24 molecular phase to another molecular phase such as *Cmca*-12. Here, through *ab initio* calculations of pressure dependent vibron frequency and direct band gap, we find that the experimental data is consistent with the *C2/c*-24 phase up to 425 GPa, and suggest that this consistency extends beyond that pressure. Specifically, we find that qualitative changes in the band structure of the *C2/c*-24 phase lead to a discontinuous drop of the direct band gap, which can explain the observed drop without a structural transition. This alternative scenario, which naturally explains the absence of hysteresis in the measurements, will hopefully motivate further experimental studies to ascertain the structure of the phase above the high pressure "phase transition".

Hydrogen was predicted to be a metal in its solid form in 1935 [1], and a superconductor with a high transition temperature in 1968 [2]. However, predicting the crystal structure of solid hydrogen proved difficult for many decades, until a list of leading candidates emerged in the 2000s [3–6]. With increasingly sophisticated computational methods and promising but limited experimental results as guides, the candidate crystal structures gradually got narrowed down to three molecular phases (*C2/c*-24, *Cmca*-12, *Cmca*-4) and one atomic phase (*I4$_1$/amd*-2) in the 2010s (number after the dash denotes the number of atoms in the unit cell) [7–16]. Computational studies have suggested that the metallization of hydrogen may occur via the closure of the band gap in the molecular phase, or a structural transition into an atomic phase. Because hydrogen nuclei are too light for the common static nuclei approximation to be accurate, predictions for the most stable phase at a given pressure and metallization pressure for a given phase depend heavily on how a particular method accounts for the zero-point motion of the nuclei. However, most recent computational studies predict the metallization to occur in the 300-500 GPa range.

Until recently, these pressures were beyond experimental reach at low temperatures. However, throughout the 2000s and the 2010s, the accessible pressure range gradually increased to reach 400 GPa, resulting first in the observation of black hydrogen at around 310-320 GPa, which is a semiconductor with a direct gap below the visible range [17–20], and second in the observation of an increase in conductivity around 350–370 GPa, where solid hydrogen potentially becomes semimetallic [21,22]. In 2017 the first experiment proposing the observation of solid metallic hydrogen at 495 GPa was reported [23,24]. A few computational studies attempted to reproduce the reported reflectance values for the atomic *I4$_1$/amd* phase [15,25], however more experiments around 500 GPa are needed to confirm this phase transition and characterize the new phase. On the other hand, a recent experiment used infrared (IR) measurements to track the vibron frequency and the direct electronic band gap of the molecular phase up to 425 GPa [26], and these measurements allow a comparison with calculations for the same range of pressures to determine the best candidates for the observed molecular phases of hydrogen.

In their experiment, Loubeyre *et al.* [26] found that at low-temperature (80 K) the IR-active vibron frequency linearly decreases with pressure from 150 GPa to 425 GPa, indicating that there is likely no phase transition in this range. They also observed that the direct band gap gradually decreases between 360–420 GPa and then abruptly drops below the minimum experimentally available value of ~0.1 eV. This is interpreted by the authors as suggesting a possible structural phase transition around 425 GPa [26]. Observations in the Raman spectra around 440 GPa were interpreted by Eremets *et al.* in a similar way [21]. However, the fact that there was an absence of hysteresis in these measurements suggests being cautious about the possibility of a structural phase transition, unless the hysteresis is too small to be observed within the limited experimental precision for pressure and wavenumber. Furthermore, by tracking the same IR-active vibron frequency, Dias *et al.* observed a first order phase transition around 360 GPa above which the original vibron mode which has a frequency of 4000 cm$_{-1}$ disappears and two IR-active modes appear at 3300 and 3000 cm$_{-1}$ [22].

Here, we investigate the vibrational and electronic structures of molecular hydrogen at a pressure range of 300 – 500 GPa using density functional theory (DFT) calculations and related theoretical techniques. We consider the three low-enthalpy structures: *C2/c*-24, *Cmca*-12 and *Cmca*-4 which have been proposed in the recent literature [7–15]. We find the enthalpies of these phases to be within 10 meV/atom of each other in this pressure range. We refer the readers who are interested in more details regarding enthalpy-based structure determination to the cited literature. Instead, here we focus on the direct comparison with the experimental data to confirm the crystal structure in the aforementioned pressure range.

We employ density functional perturbation theory (DFPT) [27,28] to calculate the zone center vibrational modes in the harmonic approximation for the pressures of 100-500 GPa. We apply the anharmonic corrections using the self-consistent phonon method as implemented in the ALAMODE package for the pressures of 400, 425, 450 and 500 GPa [29,30], using supercells to calculate the force constants[31]. The IR-active vibron frequency measured in Ref. [26] is around 4000 cm$_{-1}$ and changes linearly with pressure. The computed frequencies of the top vibron modes also behave this way in both the *C2/c*-24 and *Cmca*-12 phases. In order to distinguish between the two phases, we compute their IR activities within DFPT at 190 GPa, which is shown in **Figure 1**(a) along with the experimental results for comparison. The calculated IR absorption spectrum of the *C2/c*-24 phase agrees well with the experimental data while that of the *Cmca*-12 structure does not. In **Figure 1**(b), we present the calculated frequencies of IR-active phonon and vibron modes as functions of pressure, which closely agree with the experimental data, especially after anharmonic corrections are applied. The harmonic frequency values are also in agreement with available computational results [5,16]. Although anharmonic effects are important in hydrogen-rich systems [32,33], because of the high computational cost, we have applied them only for 400, 425, 450 and 500 GPa. The experimental data from Ref. [26] is presented by the linear fits to the observed wavenumbers, extrapolated using dashed lines. **Figure 1**(c-d) show one of the IR-active vibron modes in real space. The displacement patterns cause two out of the three molecules in a layer to vibrate with a 180° phase difference, while the third molecule vibrates much less. Further discussion of the vibrational properties can be found in the Supplementary Material. The zone center phonon frequencies of the three molecular phases as well as the atomic phase at a pressure range of 200 – 600 GPa are presented in **Figure S1** along with which modes are IR- or Raman-active.

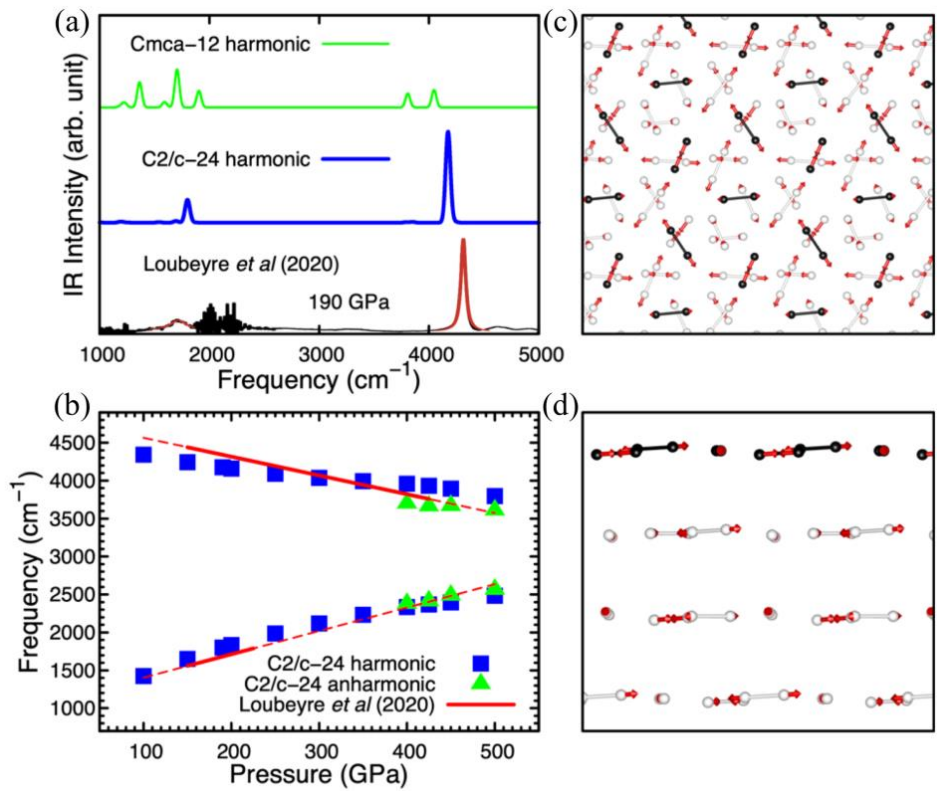

**Figure 1. Vibrational properties of the molecular phase. (a)** Theoretical simulations of the IR spectra for the *C2/c*-24 and *Cmca*-12 phases within the harmonic approximation, along with the experimentally reported IR absorption spectrum from Ref. [26]. **(b)** Wavenumbers of the IR-active phonon and vibron modes as functions of pressure. The blue squares and green triangles represent our theoretical results within the harmonic approximation and with anharmonic corrections, respectively. The red solid lines represent the linear fits of the experimental data in Ref. [26] and the red dashed lines extrapolate these fits out of the experimental range. **(c-d)** A vibron mode of the *C2/c*-24 phase is presented in real space. The red arrows represent the eigenvector of an IR-active vibron mode, and the black and white spheres represent the atoms on the top layer, and other atoms, respectively. (c) and (d) correspond to top and side views, respectively.

Based on its vibrational properties, the *C2/c*-24 phase emerges as the leading contender for the experimentally observed molecular phase up to 425 GPa. This is also the phase that has the lowest enthalpy up to 300 GPa before being overtaken by *Cmca*-12 in our calculations as well as Ref. [5]. However, the question of a structural phase transition at ~425 GPa remains unresolved, due to the fact that the vibron frequencies above this pressure are not experimentally available. To explore this question, we now turn to the other physical quantity provided in the experiment, namely the direct electronic gap ($E_{dir}$). For the electronic structure calculations, we use norm-conserving pseudopotentials generated within the Perdew–Burke–Ernzerhof generalized gradient approximation (PBE GGA) [34,35]. The calculations are conducted using the QUANTUM ESPRESSO package, and the Fermi surfaces are plotted using the XCrySDen package [36,37].

In **Figure 2**, we plot the direct gap for the three molecular phases in question for the 300 – 500 GPa pressure range. Before detailing how we arrive at these values, we first make some observations. (1) The *Cmca*-4 phase has an $E_{dir}$ of zero throughout the pressure range, which eliminates it as a candidate. (2) The *Cmca*-12 phase shows an abrupt transition from an $E_{dir}$ of ~1.7 eV to zero around 335 GPa, which makes it very unlikely to be the observed structure around that pressure, since nonzero direct gaps are observed up to ~425 GPa. (3) The *C2/c*-24 phase has a series of transitions where the band structure changes abruptly. First, a semiconductor-semimetal transition occurs at ~365 GPa (measured as an abrupt decrease in resistivity around 350–360 GPa by Eremets *et al.* [21]). In the pressure range where the $E_{dir}$ values were measured by Loubeyre *et al.* [26], we observe that the computed $E_{dir}$ values of the *C2/c*-24 phase match with their experimental counterparts up to a rigid shift of ~0.2 eV. Furthermore, the fact that the experiment did not find a direct gap above ~0.1 eV at 427 and 431 GPa is naturally explained by the fact that the computed $E_{dir}$ drops to ~0.2 eV around 405 GPa and to 0 eV around 430 GPa. In the experiment, direct band gaps are located by determining the absorption edge. However, one can also determine the direct band gaps from the absorption data using Tauc plots [38]. In **Figure 2**, we include the direct band gaps computed from Tauc plots of Loubeyre *et al.*'s raw IR data by Ranga P. Dias and Isaac F. Silvera [39]. This set of direct gaps wholly agree with our computed values in the 380–405 GPa range.

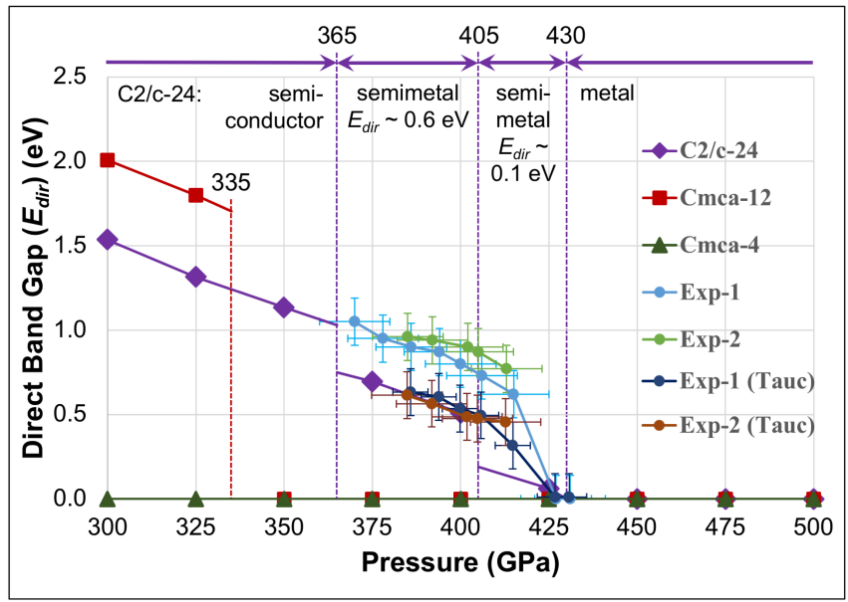

**Figure 2. Direct gap vs. pressure for molecular solid hydrogen.** The computed direct gap ($E_{dir}$) vs. pressure for the *C2/c*-24, *Cmca*-12 and *Cmca*-4 phases of solid molecular hydrogen at a pressure range of 300-500 GPa. The values obtained in Ref.[26] through IR absorption measurements are also plotted for comparison. The measurements for the rising and falling pressure are denoted by "Exp-1" and "Exp-2", respectively. The error bars are based on the reported uncertainties in the reference. On top of the plot, the four types of the *C2/c*-24 band structures and the transition pressures between them are presented. The *Cmca*-12 phase has an abrupt transition between semimetallic to metallic at 335 GPa, which is represented by a vertical red dashed line.

We note here that our computational method involves GGA which is known to underestimate band gaps. A more accurate method such as the GW approximation is expected to increase the band gaps by ~1.5 eV in this pressure range [7,10]. However, nuclear quantum effects expected to be present in these systems, which we do not take into account, are likely to lower the band gap by a similar amount in the semiconducting phase [9,13,40]. These opposite effects partially cancel out, which is likely responsible for such close agreements between the theoretical and experimental direct gap and transition pressure values. However, these effects are not expected to be fully uniform in the reciprocal space, so the experimental transition pressures may turn out to be shifted up or down with respect to these reported values. Recent works that include nuclear quantum effects have found electronic band gap values that generally agree with ours but are only provided for coarser pressure grids [40,41], and there is no established methodology to compute these effects together with the GW corrections. Therefore, by following a less computationally costly methodology that reveals the electronic structure in its full detail, we are able to study this system on a finer pressure grid, and reveal a sequence of qualitative transitions. This provides valuable physical insight into the nature of the system and the qualitative picture we present should be robust with respect to the details of the computational method.

In short, the electronic and optical behavior of solid hydrogen up to ~480 GPa observed by the two most recent experiments (Refs. [21,26]) can be explained within a single crystal structure, *i.e. C2/c*-24. Although we cannot rule out a phase transition that coincides with the direct gap closure (*Cmca*-12 also has a zero direct gap at ~425 GPa), absence of a phase transition would also explain the lack of hysteresis in Loubeyre *et al.*'s measurements. Furthermore, the fact that this phase may be stable all the way up to the transition to the atomic phase was pointed out previously in a DFT + Quantum Monte Carlo study [10], and to the best of our knowledge, no previous work has presented findings that exclude this possibility.

We now turn to the details of the band structures of the three molecular phases. We first compute their electronic structures between 300 and 500 GPa in 25 GPa increments. After a fine sampling of the first Brillouin zone, we determine the reciprocal space locations that states near the Fermi energy occupy. The next step is to visualize the Fermi surface to determine the paths on which the direct gap may reside. In **Figure 3(a-d)**, the first Brillouin zone and the Fermi surface of the *C2/c*-24 phase at 350, 400, 425 and 450 GPa are shown.

Up to 350 GPa, the overall band gap is non-zero, the Fermi surface has zero area, and the direct gap resides at the Γ-point. At 375 GPa, electron pockets around the Y-point and hole pockets around the Γ↔R′ path comprise the Fermi surface and remain so at 400 GPa. The band structure at 400 GPa is shown in **Figure 3(e)**. The direct gap for these two pressures is near the Y-point, which is shown in the left inset. The direct transition energy at the Y-point at 375 and 400 GPa is 0.72 and 0.69 eV, respectively. However, the lower of the two bands shown in the inset has a smaller effective mass along the Y→$H_3$ direction and curves up more quickly than the upper band, resulting in direct transitions with lower energy than the Y-point (0.69 and 0.51 eV for 375 and 400 GPa, respectively). We have found that that this is the direction in which the smallest direct gaps reside (**Figure S4**). By interpolating the band energies between 350 and 375 GPa, we find that the overall gap closes at ~365 GPa with the direct gap exactly at the Y-point. At this pressure, the direct gap jumps from 1.03 eV at the Γ-point to 0.76 at the Y-point, and the electronic structure becomes semimetallic.

The next transition occurs when the direct gap moves to the vicinity of the Z-point. At 400 GPa, the two doubly degenerate bands at the Z-point near the Fermi energy are both unoccupied, and they remain unoccupied while dispersing from the Z-point in all directions (T←Z→Γ is shown in the right inset of **Figure 3(e)**). At 425 GPa, these doubly degenerate bands are still unoccupied at the Z-point, but the lower-energy band divides into two bands one of which crosses the Fermi energy along some reciprocal space directions (**Figure 3(c,f)**). The direct gap is the minimum energy difference between these two bands at the k-points where this crossing occurs, which we find to be along the Z→Γ direction (**Figure S4**). By interpolating the band energies between 400 and 425 GPa, we find that this crossing first occurs at ~405 GPa when the lower-energy band becomes tangent to the Fermi energy and the direct gap changes from 0.48 eV to 0.19 eV.

As the lower of the doubly degenerate band at the Z-point drops to the Fermi level, the smallest direct transition along the Z→Γ direction gradually reduces to zero. At 450 GPa (**Figure 3(d,g)**), this doubly degenerate band at the Z-point is occupied. In the neighborhood of the Z-point, this band either stays degenerate (Z→T) or divides into two bands (Z→Γ) before crossing the Fermi energy. Therefore, by slightly perturbing the Z→T direction toward the Z→Γ direction, the direct gap between these resulting bands can be made arbitrarily small, hence the direct gap is zero. By interpolating between 425 and 450 GPa, we find that this Fermi level crossing at the Z-point happens at ~430 GPa. At all the pressures above this last transition pressure, the direct gap should remain zero.

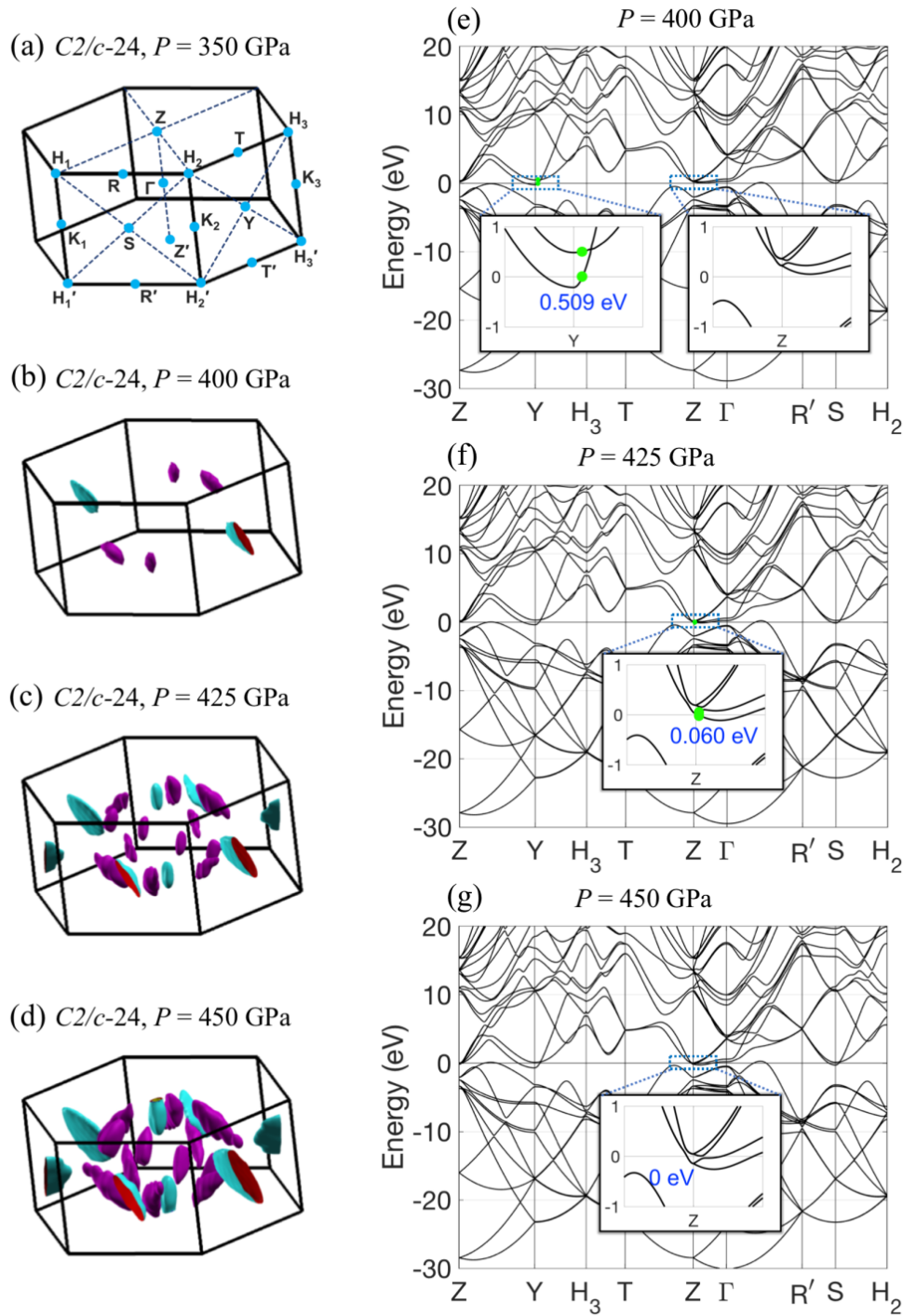

**Figure 3. Fermi surface and band structure of the *C2/c*-24 phase. (a-d)** The Fermi surface of the *C2/c*-24 phase of hydrogen at a pressure of 350, 400, 425 and 450 GPa, respectively. In (a), high-symmetry points in the reciprocal space are labeled and the Fermi surface has zero area (the material is a semiconductor at 350 GPa). **(e-g)** Band structure of the *C2/c*-24 phase of hydrogen at a pressure of 400, 425 and 450 GPa, respectively. Energies are relative to the Fermi energy. In (e), the details of the band

structure in the vicinity of the Y-point around the Fermi energy is presented as an inset. The direct gap location is denoted by the two light green spots that mark the highest occupied and the lowest unoccupied energies corresponding to the k-point of the direct gap. In (e), (f) and (g), the details of the band structure in the vicinity of the Z-point around the Fermi energy is presented as an inset. At 400 GPa (e), the two doubly degenerate bands at the Z-point near the Fermi energy are both unoccupied at and around the Z-point. At 425 GPa (f), these doubly degenerate bands are still unoccupied at the Z-point, but the lower-energy band divides into two bands one of which crosses the Fermi energy. At 450 GPa (g), one of the doubly degenerate bands at the Z-point is occupied, leading to a direct gap of zero.

The Fermi surfaces and the band structures for the *Cmca*-12 and *Cmca*-4 phases are presented in **Figure S5** along with the corresponding discussion in the Supplementary Material. All the direct gap values and reciprocal space locations for the structures we have computed are listed in **Table 1**. We note here again that GW and nuclear quantum corrections are expected to be non-uniform in the reciprocal space. For instance, Ref. [7] found that the GW corrections vary by up to ~0.6 eV in the Brillouin zone for the *Cmca*-12 phase. Similarly, the Brillouin zone locations of the *C2/c*-24 band edges found in a quantum Monte Carlo study agree with the locations we have presented, but their energies vary by up to ~0.3 eV from our values [13]. These comparisons indicate that these effects are approximately uniform in the reciprocal space but may still be large enough to slightly modify the behavior we have presented in **Figure 2**.

**Table 1. Direct band gaps of the molecular phases of solid hydrogen.** Brillouin zone location and the value of the direct band gap ($E_{dir}$) for the *C2/c*-24, *Cmca*-12 and *Cmca*-4 phases of solid molecular hydrogen at a pressure range of 300-500 GPa. When the direct gap is in the vicinity of a high-symmetry point, it is indicated as *e.g.* ~Y. When the direct gap is along a reciprocal space path connecting two points, it is indicated as *e.g.* Γ↔Y. When the direct gap is in the vicinity of a high-symmetry point along a particular direction in the reciprocal space, it is indicated as *e.g.* ~Z (Z→Γ).

| *P* (GPa) | *C2/c*-24 | | *Cmca*-12 | | *Cmca*-4 | |
|---|---|---|---|---|---|---|
| | BZ location | $E_{dir}$ (eV) | BZ location | $E_{dir}$ (eV) | BZ location | $E_{dir}$ (eV) |
| **300** | Γ | 1.54 | Γ↔Y | 2.01 | ~T | 0 |
| **325** | Γ | 1.31 | Γ↔Y | 1.80 | ~T | 0 |
| **350** | Γ | 1.13 | ~S | 0 | ~T | 0 |
| **375** | ~Y (Y→H$_3$) | 0.69 | ~S, ~R | 0 | ~T | 0 |
| **400** | ~Y (Y→H$_3$) | 0.51 | ~S, ~R | 0 | ~T | 0 |
| **425** | ~Z (Z→Γ) | 0.06 | ~S, ~R | 0 | ~T | 0 |
| **450** | ~Z | 0 | ~S, ~R | 0 | ~T | 0 |
| **475** | ~Z | 0 | ~S, ~R | 0 | ~T | 0 |
| **500** | ~Z | 0 | ~S, ~R | 0 | ~T | 0 |

In summary, among the three most likely candidates for the molecular phase of solid hydrogen, the *C2/c*-24 phase matches most closely with the recent experiments with regards to its vibron frequencies and direct gap. By using well-established and sufficiently precise methods, we have presented a detailed analysis of the band structure of this phase, which reveals that its electronic structure goes through a sequence of transitions as a result of its complicated Fermi surface, and its direct gap has a discontinuous drop to zero at ~430 GPa. This eliminates the need to postulate a structural phase transition around that pressure to explain the discontinuity in the experimental observations. Our findings suggest that the *C2/c*-24 phase may remain stable up to the transition to the atomic phase at a higher pressure, and should motivate further experimental studies to determine the crystal structure of solid hydrogen in the 400–500 GPa range.

**Acknowledgements**

This work was supported primarily by the Director, Office of Science, Office of Basic Energy Sciences, Materials Sciences and Engineering Division, of the U.S. Department of Energy under contract No. DE-AC02-05-CH11231, within the Theory of Materials program (KC2301), which supported the structure optimization and calculation of vibrational properties. Further support was provided by the NSF Grant No. DMR-1926004 which supported the determination of precise electronic structures. Computational

# Supplementary Material for "Observed Metallization of Hydrogen Interpreted as a Band Structure Effect"

Mehmet Dogan[1,2], Sehoon Oh[1,2,3], Marvin L. Cohen[1,2,(#)]

[1] Department of Physics, University of California, Berkeley, CA 94720, USA

[2] Materials Sciences Division, Lawrence Berkeley National Laboratory, Berkeley, CA 94720, USA

[(#)] To whom correspondence should be addressed: mlcohen@berkeley.edu

**Computational Methods**

We compute optimized crystal structures using density functional theory (DFT) in the Perdew–Burke–Ernzerhof generalized gradient approximation (PBE GGA) [1], using the QUANTUM ESPRESSO software package [2]. We use norm-conserving pseudopotentials with a 120 Ry plane-wave energy cutoff [3]. We use 16×16×8, 16×16×16 and 32×32×16 Monkhorst–Pack k-point meshes to sample the Brillouin zone for the *C2/c*-24, *Cmca*-12 and *Cmca*-4 phases, respectively. Self-consistent electronic energies are converged to within $10^{-8}$ Ry. All atomic coordinates are relaxed until the forces on all the atoms are less than $10^{-4}$ Ry/$a_0$ in all three Cartesian directions ($a_0$: Bohr radius). After each structure has been relaxed at a given pressure, a denser sampling of the Brillouin zone is examined to determine the band structure, using a k-point mesh that is twice as fine in all three reciprocal lattice directions. The Fermi surfaces are plotted using the XCrySDen package [4].

Using density functional perturbation theory (DFPT) [5,6] we calculate the zone center vibrational modes in the harmonic approximation for the pressures of 100-500 GPa. Convergence threshold for DFPT iterations is set to $10^{-14}$ Ry. Because of the high computational cost, we apply the anharmonic corrections using the self-consistent phonon method as implemented in the ALAMODE package only for the pressures of 400, 425, 450 and 500 GPa [7,8]. For the anharmonic corrections, we use the finite-displacement approach with a 2×2×1 real-space supercell containing 96 atoms, a 2×2×2 supercell containing 96 atoms, and a 3×3×2 supercell containing 72 atoms for *C2/c*-24, *Cmca*-12 and *Cmca*-4 phases, respectively [9]. The displacements from the equilibrium position are 0.01, 0.03, and 0.04 Å for the second, third, and fourth order terms of the interatomic force constant, respectively. We simulate the infrared (IR) spectra using DFPT when the system has a non-zero band gap, and determine the IR-active modes at other pressures based on the continuous change in the eigenmodes.

**Vibrational properties of the three molecular phases and the atomic phase**

We present the Γ-point phonon frequencies of the three molecular phases as well as the atomic *I4$_1$/amd* phase (the tetragonal 4-atom unit cell is used for this phase) at a pressure range of 200 – 600 GPa in **Figure S1**, which agree with the existing theoretical literature [10–12]. The top frequency is below and around 3000 cm$_{-1}$ for the atomic phase and the *Cmca*-4 phase, respectively, which eliminates these phases as candidate structures. The zone-center frequencies in the *C2/c*-24 and *Cmca*-12 phases are close to each other, which arises from the similarities in the atomic structures of these phases. In the figure, IR-active and Raman-active modes are marked by green and orange colors, respectively. **Figure S2** displays the relaxed crystal structures of *C2/c*-24 and *Cmca*-12 phases, both of which consist of a distorted hexagonal lattice and layers of H$_2$ molecules. The *C2/c*-24 structure exhibits ABCD stacking and molecules within a layer are not perfectly planar. The *Cmca*-12 structure exhibits AB stacking, and the molecules within a layer are perfectly planar. As seen from the similarity of their phonon frequencies, these slight differences do not significantly affect their vibrational properties.

In **Figure S3** the charge density differences caused by an IR-active and an IR-inactive vibron mode are displayed. It can be seen from the figure that although the in-plane vibration patterns are equivalent between the two modes, the relative phases of the vibrations in different layers determine their IR activities.

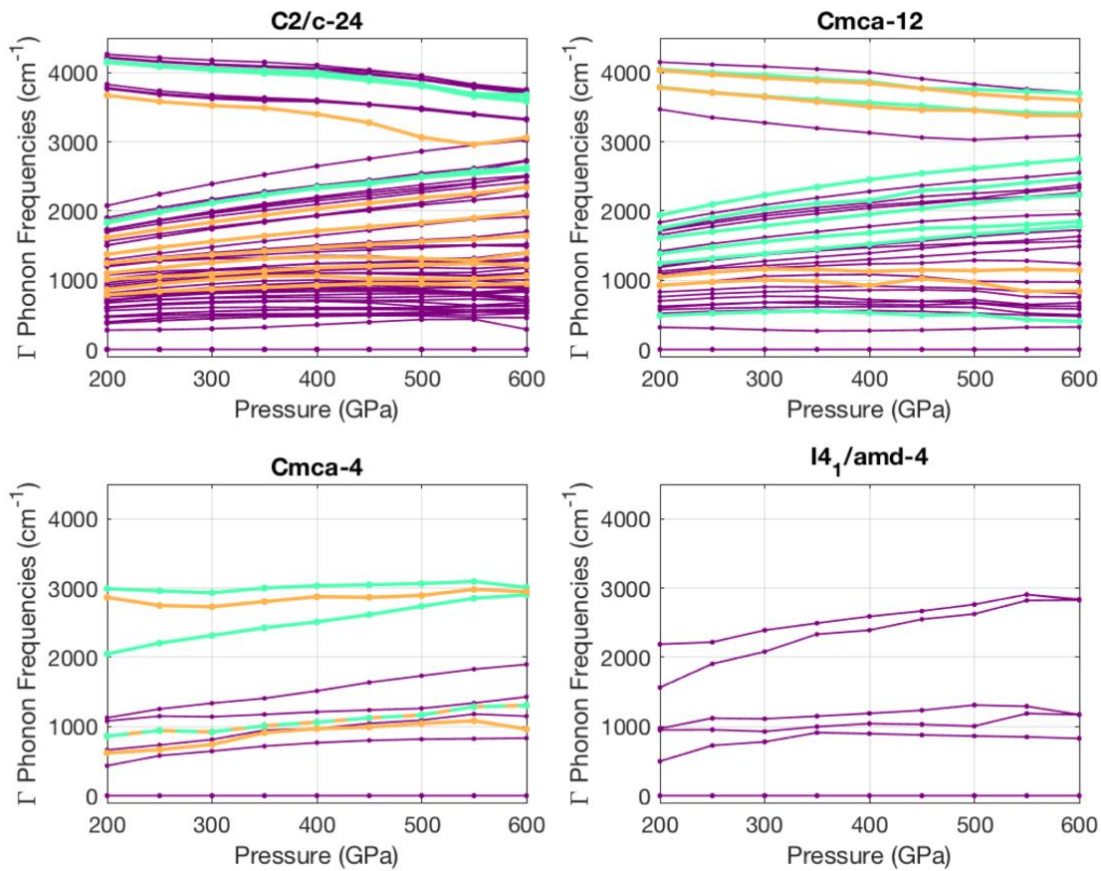

**Figure S1. Zone center phonon frequencies vs. pressure.** The Γ-point vibrational frequencies of the *C2/c*-24, *Cmca*-12, *Cmca*-4 and *I4$_1$/amd*-4 phases at 200 – 600 GPa. IR-active and Raman-active modes are highlighted using green and orange colors, respectively.

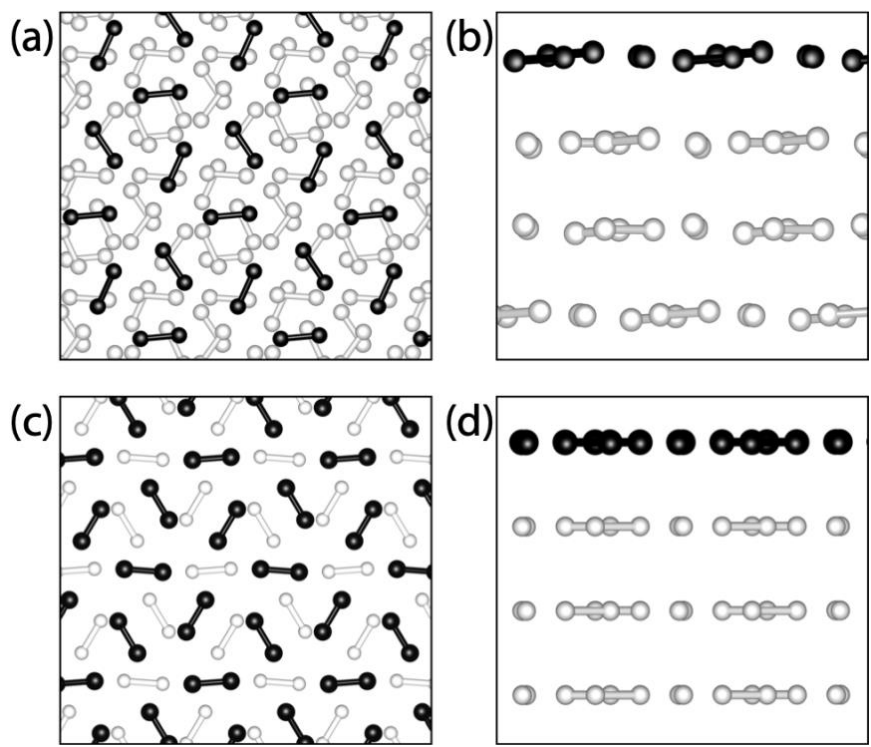

**Figure S2. Crystal structures of the *C2/c*-24 and *Cmca*-12 phases. (a-b)** The crystal structure of *C2/c*-24 phase of hydrogen at a pressure of 400 GPa. (a) Top view and (b) side view. The atoms on the same layer do not lie on one plane and the layers are stacked in ABCD stacking. **(c-d)** The crystal structure of *Cmca*-12 phase of hydrogen at a pressure of 400 GPa. (c) Top view and (d) side view. The layers are flat and stacked in AB stacking. In (a-d), the atoms on the top layer are represented by black spheres while other atoms by white spheres.

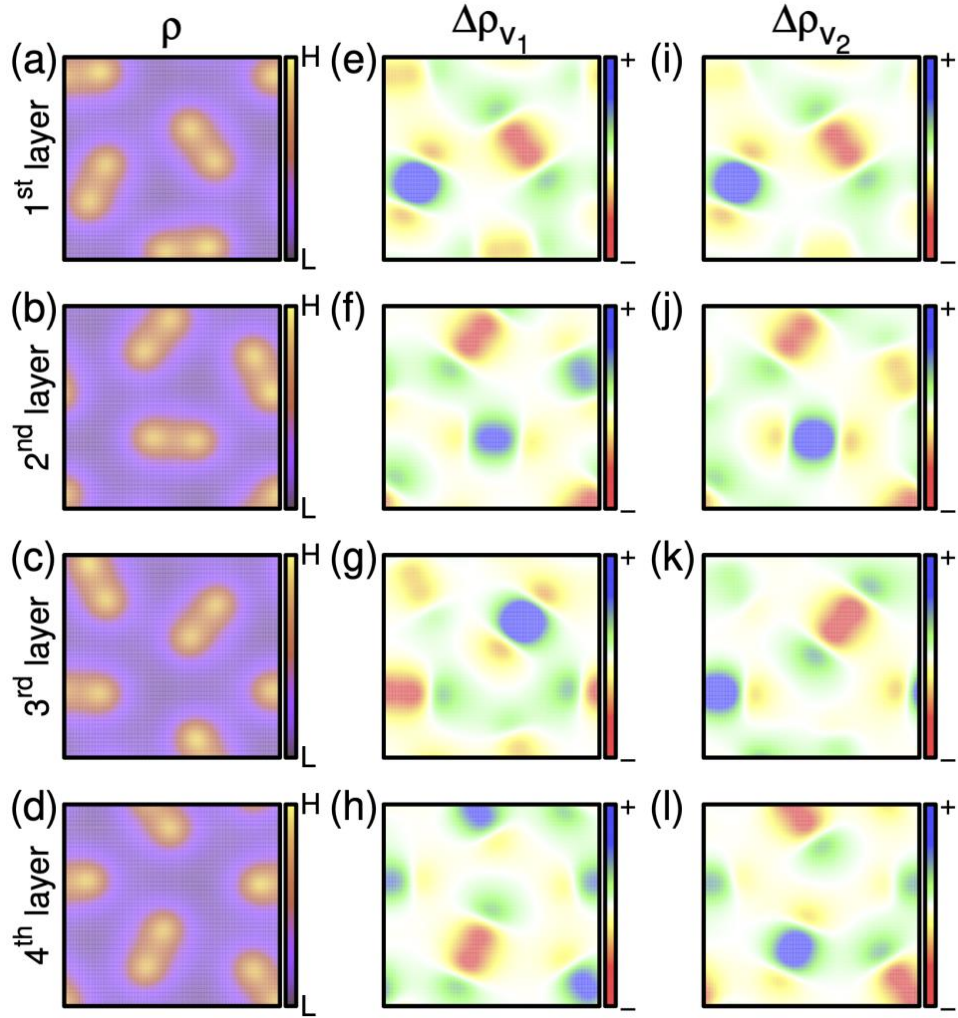

**Figure S3. Electron densities and the effects of the vibron modes in the *C2/c*-24 phase. (a-d)** Electron densities of *C2/c*-24 phase at a pressure of 150 GPa on the plane of (a) 1st (b) 2nd (c) 3rd and (d) 4th layers. The atoms are at their equilibrium positions. L and H in the sidebar correspond to low and high density, respectively. **(e-l)** Charge density differences caused by (e-h) an IR-active vibron ($v_1$) and (i-l) an IR-inactive vibron ($v_2$). The charge density differences are between the structure obtained by a perturbation from the equilibrium structure where the perturbation is equal to the normalized vibron eigenmode multiplied by 0.1 Å, and the equilibrium structure itself. Blue and red regions respectively represent increased and decreased electron density, mostly due to the shrinking and expanding molecules, respectively.

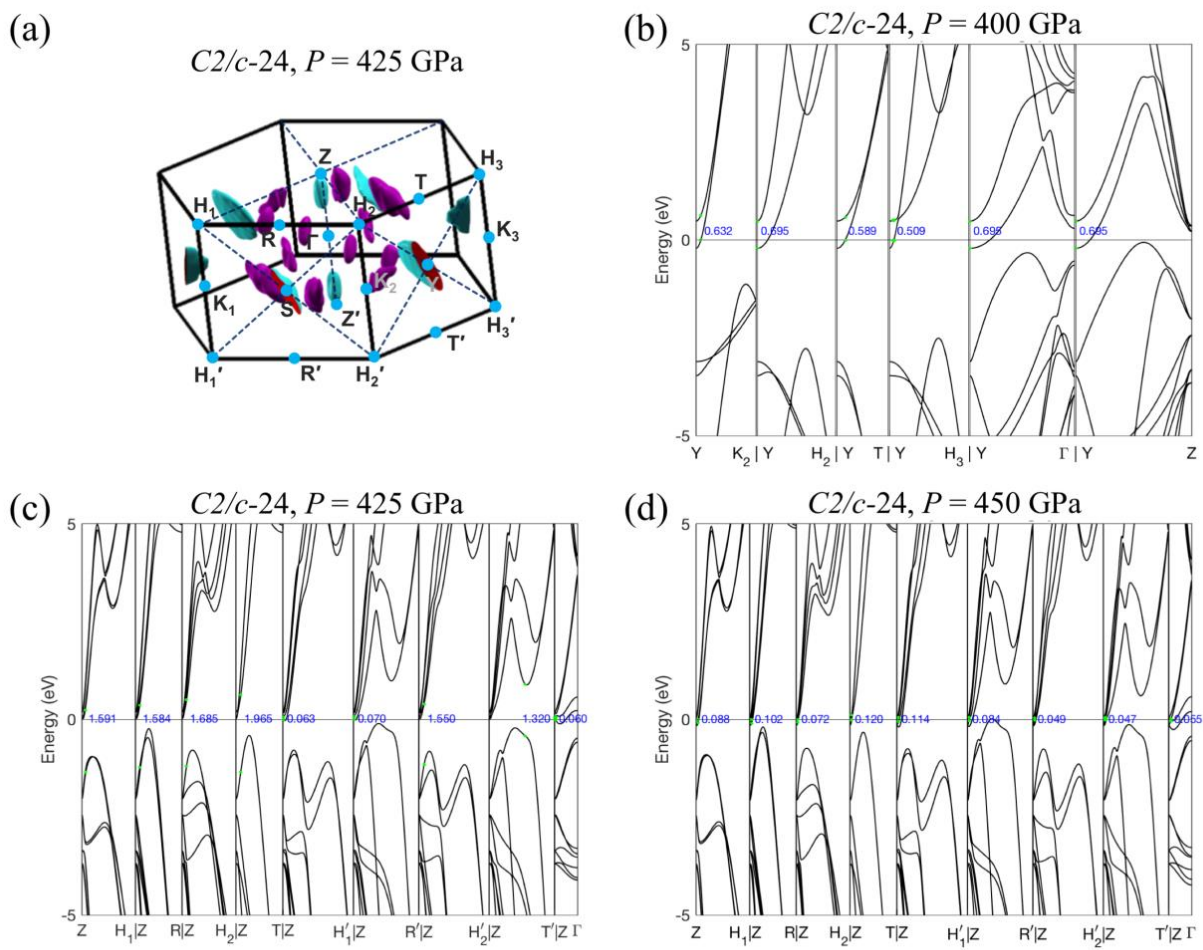

**Figure S4. Direct gap search in the reciprocal space for the C2/c-24 phase. (a)** The first Brillouin zone of the *C2/c*-24 phase of hydrogen at 425 GPa. The Fermi surface is also presented, and high-symmetry points are labeled. **(b)** Band structure of the *C2/c*-24 phase at 400 GPa along paths originating from the Y-point. For each direction, the smallest possible direct transition is denoted by the two light green spots that mark the highest occupied and the lowest unoccupied energies corresponding to transition, and the value of the transition energy is printed as a 3-decimal number in eV. Y→$H_3$ contains the smallest possible direct transition among all the paths, and thus the direct gap of the material. **(c-d)** Same as (b) but for 425 and 450 GPa, where the direct gap has shifted to the vicinity of the Z-point, thus paths originating from the Z-point are investigated. In (c), Z→Γ contains the direct gap. In (d), because the bands remain degenerate along Z→T, the direct gap can be made arbitrarily small by slight deviations from this path. The Fermi level is set to zero for all plots.

**Band Structures of the *Cmca*-12 and *Cmca*-4 phases**

The Fermi surfaces and the band structures for the *Cmca*-12 and *Cmca*-4 phases are presented in **Figure S5**. We find that the *Cmca*-12 phase becomes a semimetal at 225 GPa and remains so in all our calculations with 25 GPa increments up to 325 GPa. The direct gap resides along the Γ↔Y path right on the electron pocket which is part of the Fermi surface (see **Figure S5(a)** for 300 GPa). At 350 GPa, other electron pockets have appeared around the S-point which consist of two bands that are degenerate at the S-point (**Figure S5(b)**). These bands disperse while remaining degenerate in the S→R direction but split along the S→Γ direction. Therefore, by slightly perturbing the S→R direction toward the S→Γ direction, the direct gap can be made arbitrarily small. By interpolating the band energies between 325 and 350 GPa, we find that these bands cross the Fermi energy at the S-point at ~335 GPa, which corresponds to the metallization pressure for this phase. Further increasing the pressure results in similar doubly degenerate electron pockets to appear at the R-point at ~370 GPa, adding more area to the Fermi surface. For 425 GPa, the Fermi surface and the band structure with an inset around the S-point are presented in **Figure S5(c,f)**.

Our investigations of the *Cmca*-4 band structure have found no qualitative change in the 300 – 500 GPa range. For all these pressures, the direct transitions with the lowest energy reside around the T-point (**Figure S5(d,g)**). The doubly degenerate bands at the T-point disperse while remaining degenerate along the T→Y and the T→Z directions, but split along the directions in between, indicating that the direct gap is zero. As a final note, the *Cmca*-4 phase has a much larger density of states at the Fermi energy compared to the other two molecular phases up to 400 GPa, and does not significantly increase with pressure, therefore it is unlikely to show the IR features observed by Loubeyre *et al.* and present a large increase in conductivity at ~360 GPa observed by Eremets *et al.* [13,14] Densities of states for the structures we studied are presented in **Figure S6**.

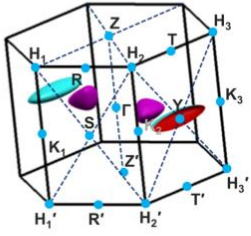
(a) *Cmca*-12, *P* = 300 GPa

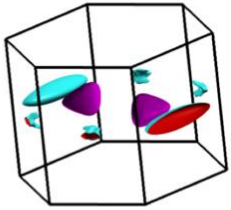
(b) *Cmca*-12, *P* = 350 GPa

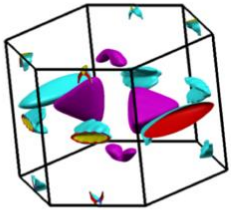
(c) *Cmca*-12, *P* = 425 GPa

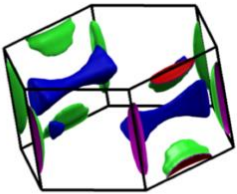
(d) *Cmca*-4, *P* = 450 GPa

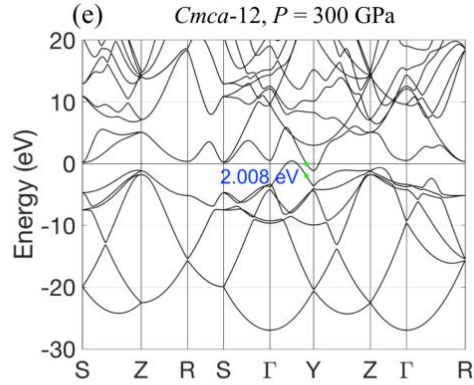
(e) *Cmca*-12, *P* = 300 GPa

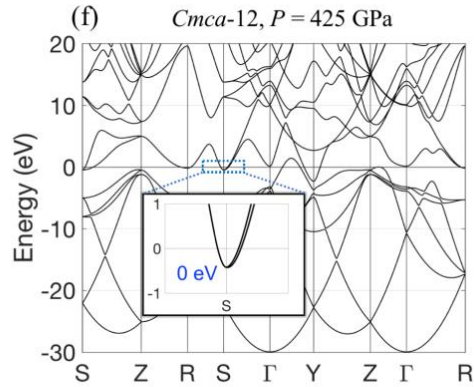
(f) *Cmca*-12, *P* = 425 GPa

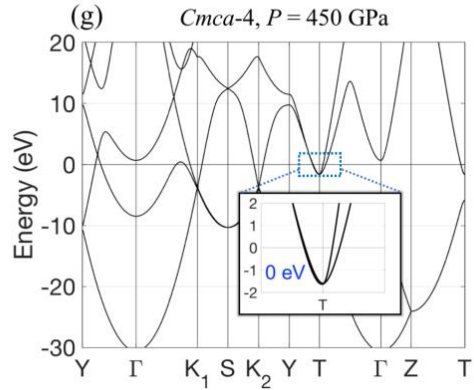
(g) *Cmca*-4, *P* = 450 GPa

**Figure S5. Fermi surface and band structure of the *Cmca*-12 and *Cmca*-4 phases.** **(a-d)** The Fermi surface of the *Cmca*-12 phase of hydrogen at a pressure of 300, 350 and 425 GPa, and of the Cmca-4 phase at a pressure of 450 GPa, respectively. In (a), high-symmetry points in the reciprocal space are labeled. **(e-g)** Band structure of the *Cmca*-12 phase of hydrogen at a pressure of 300, 350 and 425 GPa, and of the *Cmca*-4 phase at a pressure of 450 GPa, respectively. All energies are relative to the Fermi energy. For the *Cmca*-12 phase at 300 GPa (e), the direct gap resides along the Γ↔Y path and is denoted by the two light green spots that mark the highest occupied and the lowest unoccupied energies corresponding to the k-point of the direct gap. For the *Cmca*-12 phase at 425 GPa (f), a doubly degenerate band at the S-point has dipped below the Fermi energy. The details of the band structure in the vicinity of this point around the Fermi energy is presented as an inset. For the *Cmca*-4 phase at 450 GPa (g), the direct gap resides in the vicinity of the T-point. The details of the band structure in the vicinity of this point around the Fermi energy is presented as an inset.

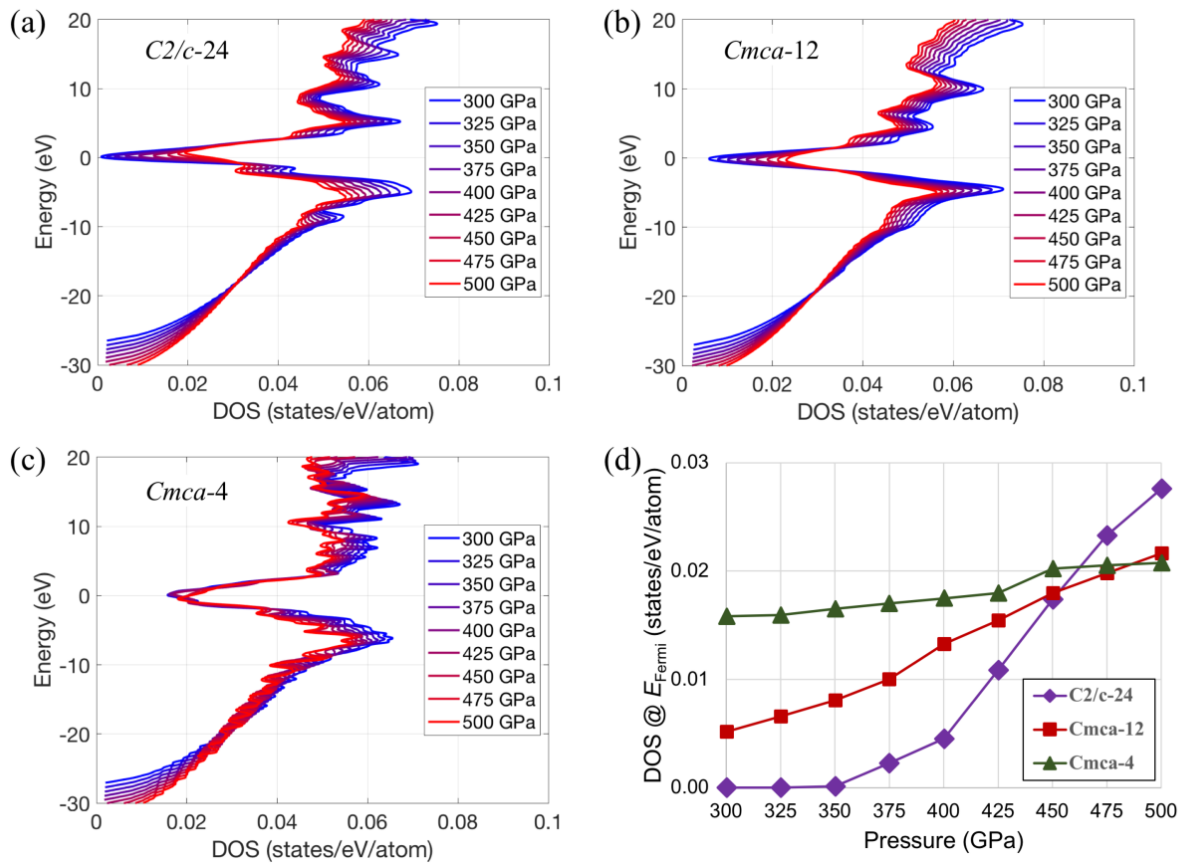

**Figure S6. Densities of states for the molecular phases.** Densities of states (DOS) for **(a)** *C2/c*-24, **(b)** *Cmca*-12 and **(c)** *Cmca*-4 phases at 300 – 500 GPa. **(d)** DOS at the Fermi energy vs. pressure for the same phases. For each case, a fine sampling of the Brillouin zone is made, and then the value is converged with respect to the decreasing Gaussian broadening used for the calculation of the DOS.